\documentclass[12pt]{article}

\usepackage{amsmath}
\begin{document}
\setcounter{page}{1}
\def\theequation{\arabic{section}.\arabic{equation}} 
\newcommand{\be}{\begin{equation}}
\newcommand{\ee}{\end{equation}}
\newcommand{\ul}{\underline}
\begin{titlepage}
\title{Is the Hawking quasilocal energy ``Newtonian''?}
\author{Valerio Faraoni \\ 
       \\ {\small Physics Department, Bishop's University} \\
{\small 2600 College Street, Sherbrooke, Qu\'ebec, Canada J1M 1Z7}\\
{\small  vfaraoni@ubishops.ca}
}
\date{} 
\maketitle
\thispagestyle{empty}

\begin{abstract}
The Misner-Sharp-Hernandez mass defined in 
general relativity and in spherical symmetry has been 
recognized as having a Newtonian character in previous 
literature. In order to better understand this aspect 
we relax spherical symmetry and we study 
the generalization of the Misner-Sharp-Hernandez 
mass to general spacetimes, {\em i.e.},  the 
Hawking quasilocal mass. The latter  is 
decomposed into a matter and a pure Weyl contribution. 
The decomposition of the Weyl tensor into an 
electric part (which has a Newtonian counterpart) and a 
magnetic one (which does not) further  splits 
the quasilocal mass into ``Newtonian'' and 
``non-Newtonian'' parts.  It is found that only  the 
electric (Newtonian) part contributes.

\end{abstract}  
\end{titlepage}   \clearpage

\section{Introduction}

The concept of mass of a gravitating system in general relativity, 
especially a non-isolated one, has been the subject of much 
research. The equivalence principle forbids the 
introduction of an energy density for the gravitational 
field because, locally, one can eliminate this 
field. The next best thing seems to be introducing a 
quasilocal energy, and this avenue has been pursued for a 
long time with the introduction of many definitions of 
quasilocal energy (see \cite{Szabados} for a review).  
Some emphasis seems to be given, in recent literature, to 
the Hawking and Hayward constructs. It is fair to say that 
a definitive prescription which is appropriate for all 
problems in relativity does not yet exist, with various 
definitions being applied to different problems and for 
different purposes. This situation is far from ideal and, 
overall, quasilocal energies remain rather abstract and 
formal concepts, at least for non-asymptotically flat 
geometries. It is only recently that the Hawking-Hayward 
quasilocal construct has been applied to more ``practical'' 
problems in cosmology, such as the Newtonian simulations of 
large scale structure formation \cite{VMA} and the 
old problem \cite{TR1, TR2, TR3, TR4, TR5, TR6, TR7, TR8, 
TR9, TR10, TR11, TR12, TR13, TR14, TR15, TR16, TR17, TR18, 
TR19} of the turnaround 
radius of the largest bound structures in the universe 
\cite{ourTurnaround}. 

The problem of deciding once and for all what is the ``mass 
of a gravitating system'' in general relativity is far from 
being solved, and the physical understanding of the various 
quasilocal definitions is the first step in this direction. 
The present manuscript contributes by examining the 
``Newtonian'' character of the Hawking mass. It has been 
pointed out \cite{BlauRollier} that, in a spherically 
symmetric spacetime, the behaviour of timelike geodesics in 
general relativity discriminates somehow between the 
Misner-Sharp-Hernandez \cite{MSH1, MSH2} and the Brown-York 
\cite{BrownYork} quasilocal energies. As seen from timelike 
geodesic observers, the Misner-Sharp-Hernandez mass (to 
which the Hawking mass reduces in spherical symmetry) plays 
the role of a Newtonian energy, while the Brown-York 
quasilocal energy plays the role of a relativistic energy 
\cite{BlauRollier}. This approach is intriguing, as it 
discloses from an unconventional but physical point of 
view, physical properties of these two quasilocal 
constructs which help understanding them better.  Here we 
want to go beyond the limitation of spherical symmetry of 
Ref.~\cite{BlauRollier} and we analyze the Hawking mass in 
general geometries. In order to make progress, one has to 
specify what is meant by ``Newtonian'' character of a 
quasilocal energy and we identify this property on the 
basis of the electric and magnetic decomposition of the 
Weyl tensor introduced long ago \cite{Matte} and widely 
used in cosmology \cite{Ellis71, C1, C2, C3, C4, C5, C6, 
Bertschinger}. With this idea in mind, it is necessary to 
relate the Hawking mass $M_\text{H}$ with the Weyl tensor. 
To this end, we first split the Hawking mass into two 
contributions, one due to matter and one to the ``pure'' 
gravitational field, {\em i.e.}, to the Weyl tensor 
$C_{abcd}$. If matter is described by a perfect fluid, the 
matter contribution to the quasilocal mass does not depend 
on the pressure.

As a second step we perform 
the splitting of the Weyl tensor into its electric and 
magnetic parts in the gravitational (Weyl) contribution to 
$M_\text{H}$ with the goal of identifying a part 
(coming from the electric part of $C_{abcd}$) which has  
a Newtonian counterpart and another part (coming from the 
magnetic part of $C_{abcd}$) which has no Newtonian 
counterpart. In so doing, we find that the gravitational 
contribution to $M_\text{H}$ is due only to the electric 
part of $C_{abcd}$ and is, in this sense,  ``Newtonian'', 
while the magnetic part gives zero contribution,  
corroborating the result  found by \cite{BlauRollier} in 
spherical symmetry (although the meaning of the adjective 
``Newtonian'' is different in our context). 

We use metric signature $-+++$, $G$ is Newton's constant, 
round (resp., square) brackets around a pair of 
indices denote symmetrization (resp., antisymmetrization), 
units in which the speed of light is unity are 
used, and otherwise we follow the notation of Wald's 
textbook \cite{Wald}.

\section{Decomposing the Hawking mass}

The Hawking-Hayward quasilocal mass contained by a 
2-surface ${\cal S}$ is defined in the following way  
\cite{Hawking, Hayward}.  Consider a spacetime $\left( 
M, g_{ab} \right)$ in general relativity and let ${\cal S}$ 
be a spacelike, 
embedded, compact, and orientable 2-surface in the 
spacetime manifold $M$. Let  $h_{ab}$ and ${\cal R}^{(h)}$ be 
the 2-metric  and Ricci scalar induced on ${\cal S}$ by 
the spacetime metric $g_{ab}$. Let $\mu$ be the volume 
2-form on ${\cal S}$ and let $A$ be the area  of 
${\cal S}$. Consider the 
congruences of ingoing ($-$) and outgoing ($+$) null  
geodesic emanating from  the surface ${\cal S}$, and let  
$\theta_{(\pm)}$ and $\sigma_{ab}^{(\pm)}$ be the expansion 
scalars and the shear tensors 
of these congruences, respectively. Let $\omega^a$ be the 
projection onto ${\cal S}$ of the commutator of 
the null normal vectors to ${\cal S}$, {\em i.e.}, the 
anoholonomicity \cite{Hayward}.  The 
Hawking-Hayward quasilocal mass is \cite{Hawking, 
Hayward}
\begin{equation}
M_\text{HH} = \frac{1}{8\pi G} \sqrt{ \frac{A}{16\pi}} 
\int_{\cal S} 
\mu \left( {\cal R}^{(h)} +\theta_{(+)} \theta_{(-)} 
-\frac{1}{2} \, \sigma_{ab}^{(+)} 
\sigma^{ab}_{(-)}    -2\omega_a\omega^a \right) \,. 
\label{HHmass}
\end{equation}
In spherical symmetry the Hawking-Hayward mass 
$M_\text{HH}$ reduces to the Misner-Sharp-Hernandez mass 
\cite{MSH1, MSH2} for a 2-sphere of symmetry 
and is a conserved Noether charge \cite{Haywardspherical}. 
The Kodama vector (defined only in general relativity in 
the presence of spherical symmetry \cite{Kodama}) is used 
in place of a 
timelike Killing vector when none exists, and generates an 
energy current (``Kodama current'') which, surprisingly, is
conserved in the absence of timelike Killing vectors 
\cite{Kodama} (the ``Kodama miracle'' \cite{miracle}). The 
Misner-Sharp-Hernandez mass is the conserved Noether charge 
corresponding to the conservation of the Kodama current 
\cite{Haywardspherical}. 

If the term $-2\omega_a \omega^a$ is dropped from 
eq.~(\ref{HHmass}), $M_\text{HH}$ reduces to the Hawking 
quasilocal prescription \cite{Hawking}, which we denote by 
$M_\text{H}$. The quantity $\omega_a \omega^a$ is 
gauge-dependent \cite{Szabados}, which is a weakness of 
the construct~(\ref{HHmass}), and we will drop it in the 
following, restricting ourselves to the Hawking mass 
$M_\text{H}$.

We are now going to decompose $M_\text{H}$ into two 
components, which can be identified as a contribution due 
to the mass-energy on the topological 2-sphere 
${\cal S}$, and one due to the gravitational field.

We take advantage of the contracted Gauss equation 
\cite{Hayward}
\begin{equation}
{\cal R}^{(h)} +\theta_{(+)} \theta_{(-)} -\frac{1}{2} \, 
\sigma_{ab}^{(+)} \sigma^{ab}_{(-)}  = h^{ac}h^{bd} 
R_{abcd} \,,\label{Gauss}
\end{equation}
where $R_{abcd}$ is the Riemann tensor, to compute the 
first three terms in the integral of 
eq.~(\ref{HHmass}). The Riemann 
tensor splits into Ricci part and Weyl part \cite{Wald}
\begin{equation}
R_{abcd}=C_{abcd} + g_{a[c}R_{d]b} -g_{b[c} R_{d]a} 
-\frac{R}{3} \, g_{a[c} g_{d]b} \,, 
\end{equation}
where $R_{ab}$ and $C_{abcd}$ are the Ricci and Weyl 
tensors, respectively, and $R\equiv {R^c}_c$ is the Ricci scalar. 
The Einstein equations in the form
\be
R_{ab} =8\pi G \left( T_{ab}
-\frac{1}{2} \, g_{ab}T \right) 
\ee
and their contraction $R=-8\pi G T$, where 
$T \equiv {T^c}_c$, yield (in conjunction with 
eq.~(\ref{Gauss}))
\begin{equation}
h^{ac} h^{bd} R_{abcd} = h^{ac} h^{bd} C_{abcd}  
 +8\pi G h^{ac} h^{bd} \left[ g_{a[c}T_{d]b} 
-g_{b[c}T_{d]a} -\frac{T}{2}  \left( g_{a[c}g_{d]b} 
-g_{b[c}g_{d]a} \right)\right] \,. \label{questa}
\end{equation}
It is easy to see that
\begin{eqnarray}
&& h^{ac} h^{bd} \left( g_{a[c}T_{d]b} -g_{b[c}T_{d]a}  
\right) = h^{ab}T_{ab} \,,\\
&&\nonumber\\
&& h^{ac} h^{bd} \left( g_{a[c}g_{d]b} -g_{b[c}g_{d]a} 
\right) =2 \,,
\end{eqnarray}
which reduces $M_\text{H}$ to the sum of a matter 
contribution and of a Weyl contribution\footnote{A version 
of eq.~(\ref{split1}) for scalar-tensor gravity appears in 
Ref.~\cite{stquasilocal}.} 
\begin{eqnarray}
&& M_\text{H} =  
\sqrt{ \frac{A}{16\pi}} \int_{{\cal S}}\mu
\left( h^{ab}T_{ab}- \frac{2T}{3}  \right)  
+\frac{1}{8\pi G} \sqrt{ 
\frac{A}{16\pi}} \int_{{\cal S}}\mu \,
h^{ac} h^{bd} C_{abcd} \,. \label{split1}
\end{eqnarray}
The first integral on the right hand side of 
eq.~(\ref{split1}), which does not contain Newton's contant $G$, 
is determined directly by the matter present on ${\cal S}$ and it
 vanishes {\em in vacuo}. The second integral, which contains $G$ 
and does not depend on matter directly,  
can be seen as a ``pure field'' contribution, although it 
contains the 2-metric $h_{ab}$ which is also determined by 
matter through the Einstein equations.

To visualize the first contribution, imagine the special 
(but very common in the literature) situation in which 
the matter content of spacetime is a perfect fluid, 
described by the stress-energy tensor 
\be
T_{ab}=\left( P+\rho \right) u_a u_b +P g_{ab} \,,
\ee
where $\rho, P$, and $u^c$ are the energy density, 
(isotropic) pressure, and 4-velocity field of the fluid, respectively. 
If we choose the 2-surface ${\cal S}$ in such a way that 
the 4-velocity $u^a$ is normal to it, {\em i.e.}, 
$h_{ab}u^b=0$, then we have 
\begin{equation}
 h^{ab}T_{ab}- \frac{2 T}{3} = 
\frac{2 \rho}{3} 
\end{equation}
and the quasilocal mass does not depend explicitly on the pressure, a 
feature which was already noted in Refs.~\cite{Hayward, 
Haywardspherical} in spherical symmetry and is now generalized to 
arbitrary spacetimes. Of course, realistically there will be an equation 
of state of the fluid relating energy density and pressure. 
Nevertheless, the property that pressures do not contribute to 
$M_\text{H}$ is noteworthy because one of the first things that one 
learns in relativity is that the pressure of a fluid gravitates together 
with its energy density, for example in the Tolman-Oppenheimer-Volkoff 
equation for interior solutions, or in the 
Einstein-Friedmann equations 
for Friedmann-Lema\^itre-Robertson-Walker cosmology \cite{Wald}. In this 
sense, it seems that this contribution to the Hawking mass behaves more 
like a Newtonian than a relativistic mass.

Let us consider now an imperfect fluid, the stress-energy tensor of 
which has the general form 
\be 
T_{ab}= \rho u_a u_b +P \gamma_{ab} +q_a u_b +q_b u_a +\Pi_{ab} \,, 
\ee 
where $\gamma_{ab}$ is the 3-metric on the 
3-space orthogonal to $u^a$ and is defined by $g_{ab}=-u_a 
u_b 
+\gamma_{ab}$, $q^a$ is a purely spatial heat current vector satisfying 
$q^c u_c=0$, and $\Pi_{ab}$ is the symmetric and trace-free shear 
tensor. For such an imperfect fluid the trace is still $T=-\rho +3P$ and 
one finds that 
\be 
h^{ab} T_{ab} -\frac{2T}{3}= \frac{2}{3}\, \rho 
+h^{ab}\Pi_{ab}= \frac{2}{3}\, \rho +{\Pi^2}_2 +{\Pi^3}_3 = 
\frac{2}{3}\, \rho -{\Pi^1}_1 \,, 
\ee 
labeling $ \left( x^2, x^3 \right)$ the coordinates on ${\cal S}$. 
Therefore, while the principal stresses associated with directions lying 
along ${\cal S}$ do not contribute to the mass $M_\text{H}$, the one 
corresponding to the third direction normal to ${\cal S}$ does 
contribute to $M_\text{H}$ (when it is non-zero) hence, to some extent, 
non-isotropic stresses gravitate according to Hawking's prescription. 

As a special case of an imperfect fluid, bulk and viscous stresses can 
be introduced as follows: 
\be
P=P_\text{(e)} +P_\text{(ne)} \,,
\ee
where $P_\text{(e)}$ is an equilibrium pressure and $P_\text{(ne)}$ is a 
non-equilibrium component, while viscosity is described by 
$P_\text{(ne)}=-\zeta \theta$, with $\eta$ a viscosity coefficient and 
$\theta =\nabla^c u_c$ the expansion of the timelike congruence with 
tangent $u^c$. Although not the most general form of an imperfect fluid, 
this is in fact the form reported in several textbooks ({\em e.g.}, 
\cite{MTW, Krasinski}) and technical articles. The 
stress-energy tensor in this case is 
\be 
T_{ab}= \rho u_a u_b +P_\text{(e)} \gamma_{ab} 
-\zeta \theta \gamma_{ab} 
+q_a u_b +q_b u_a -2\eta \sigma_{ab} \,, 
\ee 
which gives
\be
h^{ab}T_{ab}-\frac{2T}{3}= \frac{2}{3}\, \rho +2\eta {\sigma^1}_1 \,.
\ee

\section{``Newtonian'' character of the Hawking mass}

We now decompose further the gravitational contribution to 
the Hawking mass with the purpose of 
identifying its ``Newtonian'' and ``non-Newtonian'' parts 
relative to an observer with 4-velocity $u^a$. 
To give a meaning to these adjectives, we decompose the 
Weyl tensor into its electric and magnetic parts. While 
the electric part $E_{ab}$ of the Weyl tensor has a 
Newtonian analogue, its magnetic part $H_{ab}$ does not 
\cite{Ellis71} and we identify the 
``Newtonian'' contribution to 
$M_\text{H}$ with the terms due to $E_{ab}$ in 
the second integral on the right hand 
side of eq.~(\ref{split1}), and the non-Newtonian part with 
the contribution due to $H_{ab}$.

To proceed, remember that electric and magnetic parts of 
the Weyl tensor are defined relative to an observer. It is 
natural to identify the 4-velocity $u^a$ of the observer 
with the timelike unit normal to the spacelike 2-surface 
${\cal S}$. Then the electric and magnetic parts of the 
Weyl tensor are\footnote{Here we follow the definitions of 
\cite{Bertschinger}, which differ from that of 
\cite{Ellis71} in the magnetic part of the Weyl tensor and 
correct a sign error.}
\begin{eqnarray}
E_{ac}(u) &=&  C_{abcd} u^b u^d \,,\\
&&\nonumber\\
H_{ac}(u) &=&  \frac{1}{2}\, \eta_{abpq} {C^{pq}}_{ce} u^b 
u^e   \,,
\end{eqnarray}
respectively, where $\eta_{abcd}=\sqrt{-g} \, 
\epsilon_{abcd}$ with $\epsilon_{abcd}$ the alternating  
symbol and $g$ the determinant of the metric tensor 
$g_{ab}$. In other words, $\eta^{abcd} = \eta^{[abcd]}$ and 
$\eta^{0123}=1/\sqrt{-g} $. $E_{ab}$ and $H_{ab}$ are 
purely spatial, symmetric, and trace-free,
\be
E_{ab}u^a=E_{ab}u^b=
H_{ab}u^a=H_{ab}u^b=0 \,,
\ee
\be
E_{ab}=E_{(ab)} \,, \;\;\;\;\;\;
H_{ab}=H_{(ab)} \,,
\ee
\be
{E^a}_a= {H^a}_a= 0 \,.
\ee
The Weyl tensor is reconstructed from its electric and 
magnetic parts according to \cite{Ellis71, Bertschinger}
\be
C_{abcd}= \left( g_{abef} g_{cdpq} 
-\eta_{abef}\eta_{cdpq}\right) u^e u^p E^{fq} 
-\left( \eta_{abef} g_{cdpq} +g_{abef} \eta_{cdpq} \right) 
u^e u^p H^{fq} \,,
\ee
where
\be
g_{abef} \equiv g_{ae}  g_{bf} -  g_{af} g_{be} \,.
\ee
Therefore, we have
\begin{eqnarray}
C_{abcd} &=&  
\left(  g_{ae}  g_{bf} -  g_{af} g_{be}\right)
\left(  g_{cp}  g_{dq} -  g_{cq} g_{dp}\right) 
u^e u^p E^{fq}
- \eta_{abef}\eta_{cdpq} u^e u^p E^{fq} \nonumber\\
&&\nonumber\\
&\, &  -\left[ \eta_{abef}\left(  g_{cp}g_{dq} - 
g_{cq}g_{dp}\right)
+\left(  g_{ae}  g_{bf} -  g_{af} 
g_{be}\right)\eta_{cdpq}\right]u^e u^p H^{fq}\nonumber\\
&&\nonumber\\
&=& u_a u_c E_{bd} - u_a u_d E_{bc} - u_b u_c E_{ad}
+ u_b u_d E_{ac} - \eta_{abef}\eta_{cdpq} u^e u^p E^{fq}
 \nonumber\\
&&\nonumber\\
&\,& - \eta_{abef} u_c u^e H^{f}_d 
+ \eta_{abef} u^e u_d H^{f}_c
-  u_a u^p   \eta_{cdpq} H^{q}_b
+  u_b u^p   \eta_{cdpq} H^{q}_a \,.
\end{eqnarray}
By contracting twice with the (inverse) 2-metric $h^{ab}$  
most terms vanish, leaving
\be
h^{ac} h^{bd} C_{abcd} = - \eta_{abef}\eta_{cdpq}h^{ac} 
h^{bd}u^e u^p E^{fq} \,.
\ee
To summarize, after the two splittings performed, the 
Hawking mass can be written as
\begin{eqnarray}
&& M_\text{H} =  \sqrt{ \frac{A}{16\pi}} \int_{{\cal S}}\mu
\left( h^{ab}T_{ab}- \frac{2T}{3}  \right)  
- \frac{1}{8\pi G} \sqrt{ 
\frac{A}{16\pi}} \int_{{\cal S}}\mu \,
\eta_{abef}\eta_{cdpq}h^{ac} h^{bd}u^e u^p E^{fq}
\,. \nonumber\\
&& \label{summary}
\end{eqnarray}
The ``pure gravity'' contribution to $M_\text{H}$ comes 
only from the electric part of the Weyl tensor, which has a 
counterpart in Newtonian gravity \cite{Ellis71}. The 
magnetic part of $C_{abcd}$ which, on the contrary, has no 
Newtonian counterpart \cite{Ellis71}, gives zero 
contribution. In this sense, the Hawking mass is 
``Newtonian''. Although our meaning of the adjective 
``Newtonian'' is quite different from that of 
Ref.~\cite{BlauRollier}, the spirit is not too different 
and eq.~(\ref{summary}) can be seen as a statement that the 
Hawking mass is ``Newtonian'' on the same lines of the 
result of \cite{BlauRollier}. The statement is much 
stronger, in the sense that our discussion leading to 
eq.~(\ref{summary}) is not restricted to spherical 
symmetry.

\section{Conclusions}

It is intriguing that, in spherical symmetry, the Hawking 
quasilocal energy (which reduces to the 
Misner-Sharp-Hernandez one) is found to have ``Newtonian'' 
character \cite{BlauRollier}. When one tries to extend this 
result to arbitrary general-relativistic spacetimes, one 
needs to identify what ``Newtonian character'' means. While 
there are several possibilities, it is rather natural to 
think of characterizing Newtonianity by using the 
decomposition of the Weyl tensor into its electric and 
magnetic parts. In fact, the magnetic part of the Weyl 
tensor $C_{abcd}$ does not have a Newtonian counterpart, 
while its electric part does, corresponding to tidal fields 
\cite{Matte, Ellis71, Bertschinger}. The problem is how to 
relate the Hawking quasilocal mass with this decomposition 
of $C_{abcd}$. Fortunately, this question is answered 
easily by using the contracted Gauss 
equation~(\ref{Gauss}). The splitting of the Hawking mass 
$M_\text{H}$ into a matter part and a purely gravitational 
part (determined by $C_{abcd}$) is then straightforward. 
When the matter content of spacetime is a perfect fluid or 
a mixture of perfect fluids, this part of $M_\text{H}$ does 
not depend on the (isotropic) pressure, contributing to 
the interpretation of $M_\text{H}$ as a Newtonian, as 
opposed to relativistic, quantity. However, for an 
imperfect fluid $M_\text{H}$ depends on the principal 
stress in the spatial direction orthogonal to the 2-surface 
${\cal S}$ (but not on the principal stresses along the two 
directions in ${\cal S}$).

Then, the decomposition of the Weyl tensor into an electric 
(``Newtonian'') part and a magnetic (``non-Newtonian'') 
part determines a corresponding splitting of the Hawking 
quasilocal mass. However, the magnetic Weyl part vanishes 
identically in all situations in general relativity, 
leaving only the electric part that we identified as 
``Newtonian''. This means that the Hasking mass is due only 
to contributions from matter distributions and from tidal 
fields. Our procedure in identifying Newtonian and 
non-Newtonian contributions is not directly applicable to 
other quasilocal energy prescriptions. What is more, the 
characterization of ``Newtonian'' followed here may 
ultimately not be the most convenient one. These issues 
becomes more relevant in light of relativistic virial 
theorems \cite{Wiltshire} and of the application of the 
quasilocal energy to cosmological perturbations \cite{VMA, 
ourTurnaround} and will be considered further in the 
future.

\section*{Acknowledgments} 
This work is supported by the Natural Science 
and Engineering Research Council of Canada and by Bishop's University.


\clearpage
{\small \begin{thebibliography}{99}


\bibitem{Szabados} Szabados, L.B. Quasi-local 
energy-momentum and angular momentum in general 
relativity. {\em Living Rev.\ Rel.} {\bf 2009}, {\em 12}, 
4 [http://www.livingreviews.org/lrr-2004-4]. DOI: 
10.12942/lrr-2009-4   

\bibitem{VMA} Faraoni, V.; Lapierre-L\'eonard, M.; Prain. 
A. Do Newtonian large-scale structure simulations fail to 
include relativistic effects? {\em Phys. Rev. D} {\bf 
2015}, {\em 92}, 023511. DOI: 10.1103/PhysRevD.92.023511 

\bibitem{TR1} Souriau, J.M. Un mod\`ele d'univers 
confront\'e aux observations. In {\em Dynamics and 
Processes}, Proceedings of the Third Encounter in 
Mathematics and Physics, Bielefeld, Germany, Nov. 30 - Dec. 
4, 1981, Lecture notes in Mathematics vol. 1031,  
Blanchard, P.; Streit, W. eds. Springer-Verlag: Berlin, 
1981; p.~114-160.

\bibitem{TR2} Stuchlik, Z. The motion of test 
particles in black-hole backgrounds with non-zero 
cosmological constant. {\em Bull. Astronomical Institutes 
of Czechoslovakia} {\bf 1983}, {\em 34}, 129-149.

\bibitem{TR3} Stuchlik, Z.; Hledik, S. Some 
properties of the Schwarzschild-de Sitter and 
Schwarzschild-anti-de Sitter spacetimes. {\em Phys. Rev. 
D} {\bf 1999}, {\em 60}, 044006. DOI: 
10.1103/PhysRevD.60.044006 

\bibitem{TR4} Stuchlik, Z.; Slany, P; Hledik, S.  
Equilibrium configurations of perfect fluid orbiting 
Schwarzschild-de Sitter black holes. {\em Astron. 
Astrophys.} {\bf 2000}, {\em 363}, 425-439.

\bibitem{TR5} Stuchlik, Z. Influence of the 
relict cosmological constant on accretion discs. {\em Mod. 
Phys. Lett. A} {\bf 2005}, {\em 20}, 561-576.   
DOI: 10.1142/S0217732305016865 

\bibitem{TR6} Mizony, M.;  Lachi\'eze-Rey, M.  
Cosmological effects in the local static frame. {\em 
Astron. Astrophys.} {\bf 2005}, {\em 434}, 45-52.   
DOI: 10.1051/0004-6361:20042195 

\bibitem{TR7} Stuchlik, Z.; Schee, J.  
Influence of the cosmological constant on the motion of 
Magellanic Clouds in the gravitational field of Milky Way.  
{\em JCAP} {\bf 2011}, {\em 9}, 018. 
  DOI: 10.1088/1475-7516/2011/09/018 

\bibitem{TR8} Roupas, Z.; Axenides, M.; Georgiou, G.; 
Saridakis, E.N. Galaxy clusters and structure 
formation in quintessence versus phantom dark energy 
universe. {\em Phys. Rev. D} {\bf 2014}, {\em 89},  
083002.
  DOI: 10.1103/PhysRevD.89.083002 

\bibitem{TR9} Nolan, B.C. Particle and photon 
orbits in McVittie spacetimes. {\em Class. Quantum Grav.} 
{\bf 2014}, {\em 31}, 235008.
  DOI: 10.1088/0264-9381/31/23/235008 

\bibitem{TR10} Pavlidou, V.; Tomaras, T.N. Where the world 
stands still: turnaround as a strong test of $\Lambda$CDM 
cosmology.  {\em JCAP} {\bf 2014}, {\em 1409},  020. 
  DOI: 10.1088/1475-7516/2014/09/020 

\bibitem{TR11} Pavlidou, V.; Tetradis, N.; Tomaras, T.N.
Constraining dark energy through the stability of cosmic 
structures. {\em JCAP} {\bf 2014}, {\em 1405}, 017.
  DOI: 10.1088/1475-7516/2014/05/017 
    
\bibitem{TR12} Maciel, A.; Le Delliou, M.; Mimoso, J.P. 
A dual null formalism for the collapse of fluids in a 
cosmological background. arXiv:1506.07122 [gr-qc].

\bibitem{TR13} Le Delliou, M.; Mimoso, J.P.; Mena, F.C.; 
Fontanini, M.; Guariento,  D.C.; Abdalla, E. Separating 
expansion and collapse in general fluid models with heat 
flux. {\em Phys. Rev. D} {\bf 2013}, {\em 88}, 027301.  
  DOI: 10.1103/PhysRevD.88.027301 

\bibitem{TR14} Mimoso, J.P.; Le Delliou, M.; Mena, F.C. 
Local conditions separating expansion from collapse in 
spherically symmetric models with anisotropic pressures.  
{\em Phys. Rev. D} {\bf 2013}, {\em 88}, 043501. 
  DOI: 10.1103/PhysRevD.88.043501 

\bibitem{TR15} Mimoso, J.P.; Le Delliou, M.; Mena, F.C.  
Spherically symmetric models: separating expansion 
from contraction in models with anisotropic pressures.  
{\em AIP Conf. Proc.} {\bf 2011}, {\em 1458}, 487-490.  
  DOI: 10.1063/1.4734466 

\bibitem{TR16} Mimoso, J.P.; Le Delliou, M.; Mena, F.C.  
Separating expansion from contraction in spherically 
symmetric models with a perfect-fluid: Generalization of 
the Tolman-Oppenheimer-Volkoff condition and application to 
models with a cosmological constant. {\em Phys. Rev. D} 
{\bf 2010}, {\em 81}, 123514.
  DOI: 10.1103/PhysRevD.81.123514 

\bibitem{TR17} Le Delliou, M.; Mimoso, J.P.  
Separating expansion from contraction and generalizing TOV 
condition in spherically symmetric models with pressure.  
{\em AIP Conf. Proc.} {\bf 2009}, {\em 1122}, 316-319.  
  DOI: 10.1063/1.3462594 

\bibitem{TR18} Tanoglidis, D.; Pavlidou, V.; Tomaras, T.N.
Statistics of the end of turnaround-scale structure 
formation in $\Lambda$CDM cosmology. arXiv:1412.6671 
[astro-ph.CO].

\bibitem{TR19} Busha, M.T.; Adams, F.C.; Wechsler, R.H.; 
Evrard, A.E. Future evolution of structure in an 
accelerating universe. {\em  Astrophys. J.} {\bf 2003}, 
{\em 596}, 713-724.
 DOI: 10.1086/378043 

\bibitem{ourTurnaround}
Faraoni, V; Lapierre-L\'eonard, M.; 
Prain, A. Turnaround radius in an
accelerated universe with quasi-local mass.
{\em JCAP} {\bf 2015}, {\em 10}, 013.
  DOI:10.1088/1475-7516/2015/10/013

\bibitem{BlauRollier} Blau M.; Rollier, B.   
Brown-York energy and radial geodesics.  {\em Class. 
Quantum Grav.} {\bf 2008}, {\em 25} 105004.
  DOI: 10.1088/0264-9381/25/10/105004 
 
\bibitem{MSH1} Misner, C.W.; Sharp, D.H.   
Relativistic equations for adiabatic, spherically 
symmetric gravitational collapse. {\em Phys. Rev.}  {\bf 
1964}, {\em 136}, B571-B576.
  DOI: 10.1103/PhysRev.136.B571 

\bibitem{MSH2} Hernandez, W.C.;  Misner, C.W. Observer 
time as a coordinate in relativistic spherical 
hydrodynamics. {\em Astrophys. J.} {\bf 1966}, {\em 143},  
452-464. DOI: 10.1086/148525 

\bibitem{BrownYork} Brown, J.D.; York, J.W.  
Quasilocal energy and conserved charges from the 
gravitational action. {\em  Phys. Rev. D} {\bf 1993}, {\em 
47}, 1407-1419. DOI: 10.1103/PhysRevD.47.1407 

\bibitem{Matte} Matte, A. Sur des nouvelles solutions 
oscillatoires des equations de la gravitation. 
{\em Can. J. Math.} {\bf 1953}, {\em 5}, 1-16.

\bibitem{C1} Matarrese, S.;  Pantano, O.;  Saez, D. 
  A general relativistic approach to the nonlinear evolution 
of collisionless matter. {\em Phys. Rev. D} {\bf 1993}, 
{\em 47}, 1311-1323.  DOI: 10.1103/PhysRevD.47.1311 

\bibitem{C2} Bertschinger, E.;  Jain, B. Gravitational 
instability of cold matter. {\em Astrophys. J.} {\bf 1994}, 
{\em 431}, 486-494. DOI: 10.1086/174501 

\bibitem{C3} Bertschinger, E.;  Hamilton, A.J.S. Lagrangian 
evolution of the Weyl tensor. {\em Astrophys. J.} {\bf 
1994}, {\em 435}, 1-7.  DOI: 10.1086/174787 

\bibitem{C4} Bruni, M.; Matarrese, S.;  Pantano, O. 
Dynamics of silent universes. {\em Astrophys. J.} {\bf 
1995}, {\em 445}, 958-977.  DOI: 10.1086/175755 

\bibitem{C5} Croudace, K.M.; Parry, J.; Salopek, D.S.; 
Stewart, J.M. Applying the Zeldovich approximation to 
general relativity. {\em Astrophys. J.} {\bf 1994}, {\em 
423}, 22-32. DOI:  10.1086/173787 

\bibitem{C6} Kofman, L.; Pogosyan, D. {\em Astrophys. J.} 
{\bf 1995}, {\em 442}, 30-38. DOI:  10.1086/175419 

\bibitem{Hawking} Hawking, S. Gravitational radiation 
in an expanding universe. {\em J. Math. Phys.} {\bf 
1968}, {\em 9}, 598-604. DOI: 10.1063/1.1664615 

\bibitem{Hayward} Hayward, S.A. Quasilocal 
gravitational energy. {\em Phys. Rev. D} {\bf 1994} 
{\em 49}, 831-839.  
  DOI: 10.1103/PhysRevD.49.831 

\bibitem{Haywardspherical} Hayward, S.A. ``Gravitational 
energy in spherical symmetry''. {\em Phys. Rev. D} {\bf 
1996} {\em 53}, 1938-1949.
  DOI: 10.1103/PhysRevD.53.1938 
 
\bibitem{Kodama} Kodama, H. Conserved energy flux 
from the spherically symmetric system and the back reaction 
problem in the black hole evaporation. {\em Progr. Theor. 
Phys.} {\bf 1980}, {\em 63}, 1217-1228. DOI: 
10.1143/PTP.63.1217 

\bibitem{miracle} Abreu, G.; Visser, M. Kodama time: 
geometrically preferred foliations of 
spherically symmetric spacetimes. {\em Phys. Rev. D} 
{\bf 2010}, {\em 82}, 044027. DOI: 
10.1103/PhysRevD.82.044027 

\bibitem{stquasilocal} Faraoni, V. Quasilocal energy in 
modified gravity. arXiv:1508.06845.

\bibitem{Wald} Wald, R.M.  {\em General  Relativity}. 
Chicago University Press: Chicago, 1984.

\bibitem{MTW} Misner, C.W; Thorne, K.S; Wheeler, J.A. {\em Gravitation}.  
Freeman: San Francisco, 1973.

\bibitem{Krasinski}Krasi\'{n}ski, A. {\em Inhomogeneous 
Cosmological Models}. Cambridge University Press: 
Cambridge, 1997.

\bibitem{Ellis71} Ellis, G.F.R. General Relativity and 
Cosmology. In: Proceedings of the International School of 
Physics E. Fermi, Course XLVII, Varenna, Italy 1969. Sachs, 
R.K. ed. Academic Press: New York, 1971; pp.~104-182.

\bibitem{Bertschinger} Bertschinger, E. Cosmological 
dynamics: Course~1. In {\em Cosmology and Large Scale 
Structure}, Proceedings, Les Houches  
1993, Schaeffer, R; Silk, J.; Spiro, M.; Zinn-Justin, J. 
eds. Elsevier: New York, 1996; pp.~273-348 
[arXiv:astro-ph/9503125].

\bibitem{Wiltshire} Uzun, N.;  Wiltshire, D.L. Quasilocal 
energy and thermodynamic equilibrium conditions. {\em 
Class. Quantum Grav.} {\bf 2015}, {\em 32}, 16504. DOI: 
10.1088/0264-9381/32/16/165011

\end{thebibliography}}
\end{document}